\documentstyle[sprocl]{article}
\def\spose#1{\hbox to 0pt{#1\hss}}
\def\simlt{\mathrel{\spose{\lower 3pt\hbox{$\mathchar"218$}}
     \raise 2.0pt\hbox{$\mathchar"13C$}}}
\def\simgt{\mathrel{\spose{\lower 3pt\hbox{$\mathchar"218$}}
     \raise 2.0pt\hbox{$\mathchar"13E$}}}

\def\apj{{ApJ}}			
\def\apjl{{ApJ Lett.}}

\def\aap{{A\&A}}		
		
\def\aaps{{A\&AS}}

\def\mnras{{MNRAS}}

\def\pasp{{PASP}}

\def\nat{{Nature}}

\def\journal#1#2#3#4{#1, #2, #3, #4.}

\begin{document}
\title {GAMMA-RAY BURSTS:\\  CHALLENGES TO RELATIVISTIC 
ASTROPHYSICS}
\footnotetext{To appear in: Proc 18th Texas conference}

\author{MARTIN J. REES}

\address{Institute of Astronomy, 
Madingley Road, Cambridge, CB3 OHA, UK}

\maketitle
\abstracts{
Although they were discovered more than 25 years ago, gamma-ray bursts
are still a mystery.  Even their characteristic distance is highly uncertain.
All that we can be confident about is that they
involve compact objects and relativistic plasma. Current ideas and
prospects are briefly reviewed. There are, fortunately, several feasible types
of observation that could soon clarify the issues.}

\section{History}
Astrophysics is a subject where the observers generally lead, and
theorists follow behind. The topic of my talk is one where the lag is
embarrassingly large.  However, gamma-ray bursts raise issues which are
certainly fascinating to everyone involved in relativistic astrophysics.

Even though the history of gamma-ray bursts dates back more than 25
years, we still know neither where nor what they are.  The story
started in the late 1960s, when American scientists at Los Alamos had
developed a set of satellites aimed at detecting clandestine nuclear
tests in space by the associated gamma-ray emission.  Occasional
flashes, lasting a few seconds, were indeed detected. It took several
years before these were realised to be natural, rather than sinister
phenomena, and in 1973 a paper was published by Klebesadel, Strong \&
Olson entitled {\it Observations of Gamma-ray Bursts of Cosmic
Origin}.  This classic paper  reported 16 short bursts of
photons in the energy range between 0.2 and 1.5 MeV, which had been
observed during a three-year period using widely separated
spacecraft. The burst durations ranged from less than 0.1 second up to
about 30 seconds, but significant fine time-structure was observed
within the longer bursts. The bursts evidently came neither
from the Earth nor from the Sun, but little else was clear at that
time.  

    It did not take long for the theorists to become
enthusiastically engaged. At the Texas conference in December 1974,
Ruderman (1975), gave a review of models and theories. He presented a long and
exotic menu of alternatives that had already appeared in the
literature, involving supernovae, neutron stars, flare stars,
antimatter effects, relativistic dust, white holes, and some even more
bizarre  options. He noted also the tendency, still often apparent, for
theorists to ``strive strenuously to fit new phenomena into their chosen
specialities''.

 In the 1970s and 1980s, data accumulated on gamma-ray bursts, due to
a number of satellites.  Particular mention should be made of the
contributions by Mazets and his colleagues in Leningrad.  Also
important were the extended observations made by the Pioneer Venus
Orbiter (PVO).  The number of detected bursts rose faster than the
number of models -- a further index of progress is that some of the
conjectures reviewed by Ruderman were actually ruled out.

During that period, three classes of models were pursued:  those in
which the bursts were respectively in the Galactic Disc (at distances
of a few hundred parsecs), in the halo (at distances of tens of
kiloparsecs), and at cosmological distances. The characteristic
energies of each burst, according to these three hypotheses, are
respectively $10^{37}$ ergs, $10^{41}$ ergs, and $10^{51}$ ergs. The
most popular and widely-discussed option during the 1980s was that the
bursts were relatively local, probably in our Galactic Disc, and due
to magnetospheric phenomena or ``glitches'' on old neutron stars (defunct
pulsars).

It was clear that there were two statistical clues which could in
principle decide the location of gamma-ray bursts as soon as enough
data had accumulated, and selection effects were understood. One was
the number-versus-intensity of the events, which tells us whether they
are uniformly distributed in Euclidean space, or whether we are in
some sense seeing the edge of the distribution.  The other is the
degree of anisotropy.

There was already evidence that the counts of gamma-ray bursts were
flatter than the classic Euclidean slope, since otherwise more faint
bursts would have been detected by balloon experiments. This would not
of course have been unexpected if the bursts were within
the galaxy. However, the real surprise came with the launch, in April 1991,  of
the Compton Gamma Ray Observatory (GRO) satellite, whose Burst and Transient
Source Experiment (BATSE) offered systematic all-sky coverage, with good
sensitivity over the photon energy range 30 keV - 1.9 MeV.  Data from BATSE
have  transformed  the subject.

    The most remarkable BATSE result is the unambiguous evidence that
the bursts are highly isotropic over the sky. More than 1700 have now
(December 1996) been recorded, and there is still no statistical
evidence for any dipole or quadrupole anisotropy, nor for any
two-point correlation (Briggs {\it et al.\/} 1997) The lack of any
enhancement either towards the plane of the Galaxy, or towards the
Galactic Centre, is a very severe constraint on the hypothesis that
bursts come from the Galaxy. Note that they cannot be ultra-local
objects within our galactic disc: this would naturally permit
isotropy, but is ruled out by the flatness of the number counts.  The
``non-Euclidean''   counts imply that the surveys are probing to
distances where the sources are, for some reason, thinning out; the
problem is to account for this by a hypothesis that is also consistent
with the isotropy.

    The experiments on GRO have produced evidence on the spectra and
time structure of events. (For a recent review, see Fishman (1995) and
references cited therein.) Despite the large variety, there is little
doubt that gamma-ray bursts are a well-defined class of objects,
distinguished spectrally from phenomena such as X-ray bursters, and
also from the so-called ``soft gamma repeaters'' which have
substantially softer spectra.  Within this class, there are some
apparent correlations. For instance, the shorter bursts tend to be
stronger and to have somewhat harder spectra; the histogram plotting
burst durations may have two peaks; and the counts deviate most from
the Euclidean slope for the bursts with harder spectra (Kouveliotou et
al 1996).

The manifest isotropy  has tilted the balance of opinion strongly
towards a cosmological interpretation of the classical gamma-ray
bursts.  I will concentrate on discussing  the challenge posed to
theorists by that model. But I will then mention, more briefly, types of
halo model that are compatible with the isotropy since these cannot yet  be
definitively deemed irrelevant. In conclusion, I will list some
observations which might in the near future settle the issue, or at
least reduce the current level of perplexity. This talk (and the present
written version) is intended as a general overview. Fuller details, and more
extensive references, can be found in the papers from the special session on
gamma-ray bursts, elsewhere in these proceedings, or in Hartmann (1996).

\section{Models for ``Cosmological'' Bursts}

If the bursts are cosmological, then the sub-Euclidean counts imply
that the typical burst has a redshift $z$ of order 1.  The precise
redshift distribution depends on how much evolution there is in the
population. The mean redshift would be less, for instance, if the burst
rate increased with cosmic time. However, we can confidently say that
all but the very nearest of the observed bursts must have redshifts of
at least 0.2. Otherwise evolution would need to be implausibly steep
to explain the non-Euclidean counts, and nearby superclusters would
show up in the distribution over the sky. (Since the bursts exhibit such a wide
variety of time-structures, it would be astonishing if, by any measure, they
were anywhere near being standard candles. Obviously, detailed interpretations
of the counts depend on the luminosity function.)

The event rate per unit volume is very low if we are sampling a
population out to cosmological distances. It is of order $10^{-5}$ per
year per galaxy, in other words a thousand times less than the
supernova rate in galaxies.  The required energy release then amounts to
$10^{51}$ ergs in a few seconds.  (Both the estimates of
the rates and of the energy per event  would need to be adjusted in a
straightforward way, of course, if the individual events were beamed
in a small solid angle.)

\section{``The   trigger''}

    The total energy is not necessarily in itself a problem.  After
all, whenever a supernova goes off, the binding energy of a neutron
star is released in a fraction of a second, and this amounts to
$10^{53}$ ergs, a hundred times what is needed for the burst. But in a
supernova most of this energy goes to waste as neutrinos; moreover,
any impulsive electromagnetic release would not escape promptly, but
would be degraded by adiabatic expansion of the envelope before, much
later, it could leak out. So is it possible for some rare events to
occur where the energy release can escape promptly, rather than being
surrounded by an extensive opaque envelope?  The most widely favoured
possibility is coalescence of binary neutron stars (see, for example,
Narayan, Paczynski and Piran 1992).  Systems such as the famous binary
pulsar will eventually coalesce, when gravitational radiation drives
them together. The final merger, leading probably to the production of
a black hole, happens in a fraction of a second (though the swallowing
or dispersal of all the debris may take somewhat longer). The
calculated event rates for such phenomena -- and perhaps also for the
coalescence of binaries consisting of a neutron star and a black hole,
rather than two neutron stars -- are uncertain but are probably high
enough to supply the requisite rates of bursts.

\section{Fireball and gamma-ray emission}

How can the energy be transformed into some kind of fireball after
such a coalescence event? There seem to be two options. The first is
that some of the energy released as neutrinos is reconverted, when the
neutrinos collide outside the dense core where they were produced,
into electron-positron pairs or photons.  The rate of this process
depends on the square if the neutrino luminosity, and those
simulations that have so far been carried out yield rather pessimistic
estimates for the efficiency (Ruffert {\it et al.\/} 1996). The second
option is that strong magnetic fields directly convert the rotational
energy of the system into a directed outflow. This latter option
requires that the magnetic fields be amplified to strengths of order
$10^{15}$ Gauss. (Usov 1994; Thompson 1994)

The observed gamma rays seem to have a nonthermal spectrum.  Moreover,
they commonly extend to energies above 1 MeV, the pair production
threshold in the rest frame. These facts together imply that the
emitting region must be relativistically expanding. We draw this
conclusion for two reasons.  Firstly, if the region were indeed only a
light second across or less, as would be implied by the observed rapid
variability in the absence of relativistic effects, the total mass of
baryons in the region would need to be below about $10^{21}$ grams in
order that the electrons associated with the baryons should not
provide a large opacity: the rest mass energy of the baryons would
need to be 10 orders of magnitude less than that of the radiation
energy in the same volume.  Not only is this a remarkably low figure,
implying that only $10^{-12}$ of the material from the compact objects
is mixed up in the emitting region, but it would in any case imply a
relativistic expansion.  Quite apart from the baryon constraint, there
is a second reason for invoking relativistic expansion. Larger source
dimensions are required in order to avoid opacity due to photon-photon
collisions (via $\gamma + \gamma \rightarrow e^+ + e^-$).

If the emitting region is expanding relativistically, then for a given
observed variation timescale the dimension $R$ can be increased by
$\gamma^2$. The opacity to electrons and pairs is then reduced by
$\gamma^4$, and the threshold for pair production, in our frame, goes
up by $\sim \gamma$ from its ``rest'' value of $\sim 1$ MeV. A high
$\gamma$ will of course only be attained if the baryon loading is
sufficiently low, such that the ratio of total energy to rest mass
energy is larger than $\gamma$. A variety of models have been
discussed. Best-guess numbers are, for an energy of $10^{51}$ ergs, a
Lorentz factor $\gamma$ in the range $10^2$ to $10^3$, allowing
the rapidly-variable emission to occur at radii in the range $10^{14}$
to $10^{16}$ cms.  The entrained baryonic mass would need to be below
$10^{-6} M_\odot$ to allow these high relativistic expansion speeds.

 Because the emitting region must be several powers of ten larger than
the compact object that acts as  ``trigger'', there is a further
physical requirement: the original energy -- whether envisaged as an
instantaneous fireball or as a short-lived quasi-steady wind -- would,
during expansion, be transformed into bulk kinetic energy (with
associated internal cooling). It must be re-randomised and efficiently
radiated as gamma rays: this requires relativistic shocks. Impact on
an external medium (or an intense external radiation field) would
randomise half of the initial energy merely by reducing the expansion
Lorentz factor by a factor of 2. Alternatively, there may be internal
shocks within the outflow: for instance, if the Lorentz factor in an
outflowing wind varied by a factor more than 2, then the shocks that
developed when fast material overtakes slower material would be
internally relativistic (Piran 1997 and references cited therein).

In the case of expansion into an external medium, the energy would be
rethermalised after sweeping up external matter with rest mass $E/c^2
\gamma^2$ (Rees \& M\'esz\'aros 1992; M\'esz\'aros \& Rees 1993).  For
$E = 10^{51}$ ergs and $\gamma = 10^3$, only $10^{-9} M_\odot$ of
external matter need be swept up.  In an unsteady wind, if $\gamma$
were to vary on a timescale $\delta t$, internal shocks would develop
at a distance $\gamma^2 c \delta t$, and randomise most of the energy
(eg Rees \& M\'esz\'aros 1994).  For instance, if $\gamma$ ranged
between 500 and 2000, on a timescale of $\delta t$ second, internal
shocks with Lorentz factors $\sim 2$ (measured in the frame of the
mean $\gamma \simeq 1000$ outflow) would lead to efficient dissipation
at $3 \times 10^{16} \delta t$ cms.

Another important consequence of relativistic outflow is that only
material moving within an angle $\gamma^{-1}$ of the line of sight
contributes to what we observe.  Observations cannot therefore tell us
if bursts are highly beamed. Transverse pressure gradients are only
effective on angles below $\gamma^{-1}$, so material ejected in widely
differing directions behaves quite independently.  There are already a
variety of models in the literature discussing the radiation from
shocks in expanding fireballs and relativistic winds (see Piran 1997
for a recent review).  The parameters are uncertain, and the relevant
physics, involving for instance the coupling between electrons and
ions in relativistic shocks, is not sufficiently well developed to
allow accurate modelling of the radiation (see, for instance, Gallant
{\it et al.\/} 1992).  

So how is the original energy channelled from the central object into
 the outflowing fireball or wind.  Recent calculations by Ruffert {\it
 et al.\/}, 1996, suggest problems with releasing neutrino energy
 efficiently enough, and on a short enough timescale, to allow
 production of a fireball.  The options involving {\it magnetic}
 energy (cf Narayan, Paczynski \& Piran 1992) are rather less
 quantitative, but I still believe they are more promising.  As
 discussed by Usov, (1994), and Thompson (1994), a millisecond pulsar
 with a $\sim 10^{15}$ Gauss field would be slowed down in 1 second,
 its spin energy being dumped in a pair-dominated relativistic wind.
 As these authors and others have discussed, internal processes in
 such a wind could explain gamma rays with the observed spectrum and
 variability characteristics.

\section{A   ``best buy'' model} 

My personal favourite model (cf Meszaros and Rees 1997b) involves the
toroidal debris from a disrupted neutron star orbiting around a black
hole.  If this debris contains a strong magnetic field, amplified
perhaps by differential rotation, then an axial magnetically-dominated
wind may be generated along the rotation axis, perpendicular to the
plane of the torus.  The advantage of this geometry is that it seems
to offer the best chance of preventing baryon contamination, because
the baryonic material would be precluded by angular momentum from
getting near the axis without first falling into the black hole or
being on a positive-energy trajectory.

Such a configuration could arise from capture of a neutron star by a
black hole of less than $5M_\odot$, this mass limit being required
because otherwise the neutron star would be swallowed before
disruption.  Alternatively, it could be the outcome of the merger of
two neutron stars, where most of the mass collapses to a black hole,
leaving some fraction of the original material in orbit around it. (cf
Ruffert {\it et al.\/} 1996; Jaroszynski 1996)

The available energy in this model is the kinetic or gravitational
energy of the neutron-star debris left behind in the torus, plus the
spin energy of the hole itself (which, being the outcome of binary
coalescence, is almost guaranteed to have a high angular
momentum). Near the axis, we would expect maximal dissipation (from
fields threading the hole or anchored in the torus) but minimum
baryonic loading. The Lorentz factor would therefore be largest along
the axis. Indeed, a narrow channel, essentially free of baryons, may
carry a Poynting-dominated outflow, energised by the hole via the
Blandford-Znajek process.

Along any given line of sight, the time-structure would be determined
partly by the advance of jet material into the external medium, but
probably even more by internal shocks within the jet, which themselves
depend on the evolution and instabilities of the torus, from its
formation to its eventual swallowing or dispersal.  Even if the bursts
were caused by a completely standardised set of objects, their
appearance would be likely to depend drastically on orientation
relative to the line of sight.  Other phenomena as yet undiscovered --
for instance some new class of X-ray or optical transient -- may be
attributable to gamma-burst sources viewed from oblique orientations.

\section{Physics of the emission mechanism}

     We are a long way from a convincing model for what triggers 
gamma-ray bursts: coalescing compact binaries seem likely to be implicated,
but we should remain open-minded to more exotic options.  A precise
description of the dynamics, along with the baryon content, magnetic
field, and Lorentz factor of the outflow, might allow us to predict
the gross time-structure. But even then we could not predict the
intensity or spectrum of the gamma rays -- still less answer key
questions about the emission in other wavebands -- without also having
an adequate theory for particle acceleration in relativistic
shocks. We need the answers to the following poorly-understood
questions:

(i) Do relativistic shocks yield particle spectra that obey power
laws? This is in itself uncertain: the answer probably depends on the
ion/positron ratio, and on the relative orientation of the shock front
and the magnetic field (e.g. Gallant {\it et al.\/} 1992).

(ii) In ion-electron plasmas, what fraction of the energy goes into
the electrons?

(iii) Even if the shocked particles establish a power law, there must
be a low-energy break in the spectrum at an energy that is in itself
relativistic.  But will this energy, for the electrons, be $\Gamma_s
m_e c^2$, or $\Gamma_s m_p c^2$ (or even, if the positive charges are
heavy ions like Fe, $\Gamma_s m_{\rm Fe} c^2$?

(iv) Can ions be accelerated up to the theoretical maximum where the
gyroradius becomes the scale of the system? If so, the burst events
could be the origin of the highest energy cosmic rays (an interesting
possibility addressed by other speakers at this conference)

(v) Do magnetic fields get amplified in shocks? This is relevant to
the magnetic field in the swept-up external matter outside the contact
discontinuity, and determines how sharp the external shock actually is
(cf Mitra 1996)

(vi) Can radio emission be generated by a coherent process? If not,
the usual surface brightness constraint implies that there would be
little chance of detecting a radio   ``afterglow''.

These questions, crucial for gamma ray bursts, are  also relevant to other
phenomena. For example, Lorentz factors of at least 10 (and probably
electron-positron plasmas) exist in the compact components of strong
extragalactic radio sources probed by VLBI.

If one is prepared to parametrise the uncertainties implicit in the
above questions, predictions can be made of how the spectrum would
evolve during a burst with simple time-structure. (eg Meszaros {\it et
al.\/} 1994, Tavani 1996, Meszaros and Rees 1997).  For a wide range
of parameters, the associated X-rays would be above the threshold of
small omnidirectional detectors such as those developed for the High
Energy Transient Explorer (HETE) It was therefore a real setback to
the subject -- particularly to the prospect of using concurrent X-ray
or UV emission to pinpoint the burst locations more accurately -- when
HETE failed to go properly into orbit.

    After the main emission is over, the fireball material would
continue to expand, with steadily-falling Lorentz factor, into the
external medium.  Associated optical emission may persist for hours or
even days. This is long enough to allow an initial detection with
BACODINE to be followed up by raster scans with a 1 m telescope, that
could detect even emission down to 15th magnitude.

\section{``Extended  Halo'' Models}

In the interests of balance, I would like to make a few remarks about
the alternative idea that the bursts are not from cosmological
distances, but instead come from within our own galaxy.  Classical
gamma-ray bursts could be isotropic enough to be consistent with the
BATSE data if they came from neutron stars ejected from our Galactic
Disc at more than 700 km s$^{-1}$, which remained active, bursting
sporadically, for long enough to allow them to reach distances of at
least 100 kiloparsecs. They may either escape from the galaxy, or be
on very extended bound orbits. These high velocity objects could be a
special subset of pulsars.  The typical velocities of the pulsars
sampled in surveys may be as high as 400 km/sec (Lyne and Lorimer
1994, J. Taylor, these proceedings). Moreover, those that formed with
higher kick velocities and/or with strong magnetic fields (and
therefore short lifetimes) are under-represented in surveys; we cannot
exclude the possibility that a high fraction of newly-formed pulsars
are of such types.  If we conservatively suppose that they are only a
few percent of all pulsars, and form in our Galactic Disc at a rate of
about 1 per thousand years, then each must produce $10^6$ bursts, of
typical energy $10^{41}$ ergs. (If the relevant objects formed at a
rate of one per 100 years, the requirements placed on each would be
ten times more modest.) Repetition would not necessarily be expected,
since each neutron star could in principle continue bursting at a slow
rate for more than a billion years. However, if the bursts came in
groups, rather than being independent poissonian events, repetition
would not be impossible.

  \section{  Fitting the isotropy}

   Podsiadlowski has done detailed calculations of whether such a
population can provide an isotropic distribution. He shows this is
indeed possible for long-lived bursters whose orbits take them out
beyond 100 kiloparsecs. An important feature of such orbits is that,
because the galactic halo potential is not spherical (and may indeed
be rather irregular at such large distances) objects do not conserve
their angular momentum and therefore, even if they started off near
the centre of our Galaxy, they need not return so close to the centre
in later orbits. This effect helps to ensure greater isotropy.
Another possibility, favoured by Lamb (1995), is that the typical
objects have velocities above a thousand kilometres per second, and
are escaping the Galaxy completely.  In this case, the best fit is
obtained if the bursts do not start until after a delay of around
$10^7$ years, by which time all neutron stars have reached distances
of 30 kiloparsecs or more.

I think it is fair to say that such models need to be carefully tuned
in order to fit the existing isotropy data, but that, though perhaps
unappealing, they cannot be ruled out. The constraints on orbital
parameters would be eased in alternative schemes where the neutron
stars {\it formed} far out in the halo (being perhaps, as Woosley
(1993) has discussed, relics of an early population of halo stars)
rather than being ejected from the disc.

\section{Mechanisms for halo bursts}

 If a   ``halo''   model is to be taken seriously, there must be an
acceptable mechanism for producing the succession of $10^{41}$ erg
bursts, spread over a very long timescale. Two options have been
proposed (Podsiadlowski, Rees \& Ruderman 1995).

The first possibility is that the relevant subset of neutron stars
start off with a super-strong ($\simgt 10^{15}$ Gauss) magnetic field.
This field, penetrating the core of the neutron star, would gradually
rise towards the surface through buoyancy effects, thereby causing
stress in the crust.  The timescale for the buoyancy is estimated to
be at least $10^6$ years.  Acceptable models require that it be
$\simgt 10^9$ years.  The total stored energy is $\sim 10^{47}
(B/10^{15} G)$ ergs. The energy depends linearly on $B$, rather than
quadratically, because the field in the core is concentrated into
tubes where its strength has a standard value of $\sim 3 \times
10^{15} G$.

The crust gets stretched as the field drifts outwards.  The units in
which energy is released depend on how much stress can build up in the
crust, and what fraction is released when the crust cracks. This is
a complicated problem in asteroseismology. However, a release of
$10^{41}$ ergs per event is plausible, in which case the total stored
magnetic energy would be sufficient to supply the requisite $10^6$
events.

The second very different option for triggering halo bursts involves
asteroidal impacts on to a neutron star.  Each event requires, on
energetic grounds, the impact of $10^{21}$ grams.  The main problem
with this idea is that such asteroidal or cometary bodies would be
tidally disrupted too far out to give a sudden enough event.  A
possible solution is that the debris from the disrupted body squashes
down the magnetic field, which then rebounds, generating high electric
fields and thereby a pair cascade.  Alternatively, the debris may form
a disc which accumulates before triggering a sudden electromagnetic
release when it couples its rotation to that of the neutron star.

The total impacting mass, to get enough bursts per star, must be
$10^{27}$ grams.  It is not impossible (especially now we know that
planetary systems can exist around pulsars) that a neutron star could
carry with it $\simgt 10^{27}$ gm of asteroidal debris.  However, a
larger reservoir, plus at least one large planet, is needed in order
for enough of these planetesimals to be perturbed on to near radial
orbits. We know that at least one pulsar has a planetary system. This
fact, plus the evidence that even typical pulsars may have velocities
of 400 km s$^{-1}$, suggests that models of this kind should not be
dismissed.  Whatever the bursts turn out to be, the primary trigger,
and the efficient conversion of its energy into gamma rays, involve
physical conditions that are extreme and unfamiliar.

\section{How can we settle the debate?}

There is no convincing and fully worked out model for the bursts on
either the halo or the cosmological hypothesis. Neither option,
however, seems to violate any cherished beliefs in physics or
relativistic astrophysics.  The issue is one of plausibility, and how
one weighs different lines of evidence.  The isotropy would be a
natural consequence of the cosmological hypothesis.  But the level of
isotropy so far revealed by BATSE, which restricts any dipole or
quadrupole anisotropy below the few per cent level and shows no
evidence for clumping on smaller scales, could be accommodated in a
halo hypothesis if high speed neutron stars were implicated.

In April 1995, the 75th anniversary of the Shapley/Curtis debate,
there was an interesting debate in Washington on the location of
gamma-ray bursts -- a current issue offering some amusing parallels to
the earlier controversy concerning the distances of the nebulae.  The
two main protagonists were Don Lamb and Bohdan Paczynski (a written
version of the argument appears in Lamb (1995) and Paczynski
(1995)). I had the privilege of acting as the moderator in this
debate, perhaps because I was one of the few people who had not
already taken a firm stance on one side of the issue or the
other. There was an agreement among all participants that the issue
would be settled only by more data. Indeed, there was a broad
consensus on some particular tests that could be crucial, or at least
highly suggestive.  Among these might be the following.

  Most valuable of all would be a firm identification of a burster
with some other class of object. The stumbling block here is the poor
positional accuracy of most gamma-ray detectors.  BATSE itself has
error circles of 1 or 2 degrees for the brightest bursts, and more
than 5 degrees for the fainter ones. However, the locations of some
bursts have been pinned down with a precision of minutes of arc or
better by triangulation experiments involving deep space probes; this
technique utilises the rapid time structure, which, when recorded and
timed by detectors separated by 10 light minutes or more, allows
accurate positioning. There is still no firm identification of any
classical gamma-ray burst, though there are tantalising indications
that some of the brighter bursts may be correlated with galaxies or
clusters of galaxies, whose distances are not inconsistent with what
is expected on the cosmological hypothesis. (It is disappointing,
incidentally, that the failure of the recent Mars probe, which would
have carried a small gamma-ray detector, means that we now lack the
requisite deep-space network for obtaining accurate   ``triangulation''.)

Even though the gamma-ray positional information is poor, one might be
able to pin down the position of the sources more accurately if they
displayed concurrent transient emission in some other
waveband. Various projects have been undertaken in the optical and
radio band. Ground-based observers can be notified of a BATSE event
within a few seconds; a small telescope can then be rapidly slewed to
seek an optical counterpart within less than a minute. No such
counterparts have been detected, nor have radio searches yet yielded
positional or timing coincidences. The likely strength of gamma-ray
bursts in the optical or radio band is uncertain and highly
model-dependent. Indeed, any detection in these wavebands would have
the bonus that it would help to narrow down the range of possible
models and emission mechanisms.  However, most theories predict that
there should be substantial spectral extension from gamma-rays down
towards the X-rays, so it would seem less of a gamble to seek X-ray
counterparts.

   If the bursts have a local rather than cosmological origin, then,
at some level, anisotropies over the sky would be bound to show up.  A
particularly crucial test would be feasible if bursts more than ten
times fainter than those recorded by BATSE could be detected.  It
would then, according to the halo hypothesis, be feasible to detect
bursts from the halo of Andromeda, and there should be a definite
excess of weak events from that direction (Bulik \& Lamb 1997,
Ruszkowski \& Wijers 1997). The lack of such a trend would severely
embarrass halo models. A specific proposal has been made to look at a
10 degree field around Andromeda with 20 times the sensitivity of
BATSE. But X-ray detectors are more readily available and more
sensitive. For this reason, and also because the x-ray emission from
bursts seems stronger than a straight extrapolation of the gamma-ray
spectrum suggests (Preece {\it et al.\/} 1996), the best prospects for
testing the halo model might be from long-duration observations of
Andromeda and other nearby galaxies.

The cosmological interpretation of bursts would be confirmed, as
Paczynski (1986) first pointed out, if there were evidence of
gravitational lensing by an intervening galaxy.  If a suitable galaxy
lay along the line of sight to a cosmologically distant burst,
radiation would reach us by two or more different paths, whose light
travel times would differ typically by weeks or months. We would
therefore detect two bursts from the same direction. Even though the
positions could not be pinned down accurately, the elaborate time
structure of each burst is highly distinctive, and if two
bursts with identical ``fingerprints'' were detected from within the
same error circle, this would be compelling evidence that they were
actually separate gravitationally-lensed images of the same burst.
(As a technical point, it should be noted that microlensing by stars
or substellar objects would only introduce differences on millisecond
timescales between the two burst profiles (Williams and Wijers 1997),
and therefore would not vitiate this possibility.)

Unfortunately, the probability that a galaxy lies along a random line
of sight to a high redshift object is below one per cent, the exact
value depending of course on the presumed redshift of the
burst. Moreover, because BATSE can only observe a given direction in
the sky for about 40 per cent of the time it is more likely than not
that, if a lensed event occurred, the recurrence would be missed
because it would occur during dead time.  Taking these effects into
account, it is rather marginal whether we would expect BATSE to detect
a single instance of this lensing before it dies, even if the bursts
indeed come from cosmological distances.  However, if we were lucky,
such a double burst could clinch the cosmological interpretation.

A further issue which has figured strongly in the debate on the
location of gamma-ray bursts concerns the existence or otherwise of
spectral features attributable to cyclotron lines.  This is a
technical controversy which I will not enter here. However, its
relevance lies in the fact that halo models involve neutron stars,
where the magnetic fields are expected to be in the range such that
cyclotron lines should be in the hard X-ray band.  On the other hand,
the fields in the emitting regions of cosmological fireballs or
relativistic winds would not, even when relativistic effects are taken
into account, give rise to such features.

The controversies in the Shapley-Curtis debate were settled within a
few years. Our knowledge of extragalactic astronomy thereby made a
forward leap, and astronomers moved on to address more detailed
issues.  I'm enough of an optimist to believe that it will be only a
few years before we know where and perhaps even what, the
gamma-bursters are. Even if this optimism is misplaced, I am
completely sure that these mysterious phenomena will serve as a
continuing challenge and stimulus to theorists, and will remain high
the agenda of future Texas Conferences.

I thank my colleagues Josh Bloom, Peter Meszaros, Philipp
Podsiadlowski, and Ralph Wijers for collaboration and discussion. I am
also very grateful to the members of the BATSE team, especially Jerry
Fishman and Jim Brainerd, for updating me on the observations and
answering several queries.

\vfill
\end{document}